\documentclass[letterpaper, 10 pt, conference]{ieeeconf}  % Comment this line out if you need a4paper
\pdfoutput=1

\IEEEoverridecommandlockouts                              % This command is only needed if 
\usepackage{graphicx} % for pdf, bitmapped graphics files
\usepackage{amsmath} % assumes amsmath package installed
\usepackage{amssymb}  % assumes amsmath package installed
\usepackage{amsfonts}
\usepackage[hidelinks]{hyperref}
\usepackage[caption=false]{subfig}
\usepackage{booktabs}
\usepackage{multirow}
\usepackage{cite}
\graphicspath{{./figures/}}

\title{ \bf
Particle-Filter-Enabled Real-Time Sensor Fault Detection Without a Model of Faults
}

\author{Matthew A. Wright and Roberto Horowitz% <-this % stops a space
\thanks{This work was supported by the California Department of Transportation and the Anselmo Macchi Fellowship.}% <-this % stops a space
\thanks{Mechanical Engineering Department and Partners for Advanced Transportation Technologies,
	University of California, Berkeley, CA 94720, USA
        {\tt\small \{mwright, horowitz\}@berkeley.edu}}%
}

\begin{document}

\maketitle
\thispagestyle{empty}
\pagestyle{empty}

%%%%%%%%%%%%%%%%%%%%%%%%%%%%%%%%%%%%%%%%%%%%%%%%%%%%%%%%%%%%%%%%%%%%%%%%%%%%%%%%
\begin{abstract}
We are experiencing an explosion in the amount of sensors measuring our activities and the world around us.
These sensors are spread throughout the built environment and can help us perform state estimation and control of related systems, but they are often built and/or maintained by third parties or system users.
As a result, by outsourcing system measurement to third parties, the controller must accept their measurements without being able to directly verify the sensors' correct operation.
Instead, detection and rejection of measurements from faulty sensors must be done with the raw data only.
Towards this goal, we present a method of detecting possibly faulty behavior of sensors.
The method does not require that the control designer have any model of faulty sensor behavior.
As we discuss, it turns out that the widely-used particle filter state estimation algorithm provides the ingredients necessary for a hypothesis test against all ranges of correct operating behavior, obviating the need for a fault model to compare measurements.
We demonstrate the applicability of our method by showing its ability to reject faulty measurements and accuracy in state estimation of a nonlinear vehicle traffic model, without information of generated faulty measurements' characteristics.
In our test, we correctly identify nearly 90\% of measurements as faulty or non-faulty without having any fault model.
This leads to only a 3\% increase in state estimation error over a theoretical 100\%-accurate fault detector.
\end{abstract}

%%%%%%%%%%%%%%%%%%%%%%%%%%%%%%%%%%%%%%%%%%%%%%%%%%%%%%%%%%%%%%%%%%%%%%%%%%%%%%%%
\section{Introduction}

Much ado has been made about the contemporary explosion of ubiquitous sensors and actuators in engineered artifacts and the built environment.
``Big data,'' the ``Internet of Things'' - these and other recently-coined, oft-heard phrases call back to the idea that, in the past few years, a sea change has occurred in our ability to measure the world.
For control and systems engineers, these new data increase the ability to measure and precisely control built-environment systems, with applications like localized responsive building HVAC regulation, connected vehicles for traffic management and accident reduction,
and power systems, up to and including automated coordination between power producers, consumers, and storers in a ``smart'' grid.
When third-party measurements are collected and submitted by end users of Internet-of-Things devices and the like, fusion of these data can enhance the estimation and control of these sorts of highly-sensed and/or highly-actuated systems.

However, these new data can present their own problems.
Like any sensor, these third-party sensors cannot always behave perfectly, and data may be collected and submitted by a sensor that is faulty, overly noisy, tampered-with, or otherwise misreporting.
This problem of sensor fault detection is a problem as widespread and highly-studied across industries as the use of sensors themselves.
Some fault detection methods attempt general applicability to large classes of systems, while others take advantage of application-specific knowledge to detect failures along known failure modes.
See \cite{gao_survey_2015} for a recent survey that covers both general and specific methods in the literature.
For the example application considered in this work, that of vehicle speed measurements for traffic control, we show a general algorithm.
Since third-party sensors in this example vary widely, it was necessary to focus on a generally-applicable scheme.

In particular, we make use of the particle filter, a widely-used state estimator for general stochastic nonlinear systems \cite{doucet2011tutorial}.
We are not the first authors to use a particle filter state estimator for sensor fault detection purposes.
The particle filter is a flexible and widely-applicable state estimator, and it has been used for fault detection by providing ``analytical redundancy'' \cite{gao_survey_2015}.
This means its estimate is used to produce a predicted sensor reading that is compared to received measurements, with any difference being potential evidence of faults.
Often, however, it is assumed that the engineer has models for sensor faults, and the particle filter estimate is used to select a maximum-likelihood estimate among the known types of sensor modes (cf. interacting-multiple-model estimators, and e.g., \cite{wei_adaptive_2009, de_freitas_rao-blackwellised_2002, caron_particle_2007, tadic_particle_2014}).
Some recent works have considered a more probabilistic handling of potential faults, in keeping with the particle filter's nature as a probabilistic state estimator.
In \cite{tadic_particle_2014}, for example, when a sensor measurement diverges from the particle filter estimate, the errors between the observed and predicted measurements over the next few timesteps are collected.
After a while, a hypothesis test is performed to determine whether the errors are large and consistent enough to confirm a sensor fault, and which (if any) known fault models are active.

Two common threads among these particle-filter-based methods are assumptions of knowledge of fault models and that the sensors will repeatedly report so that potential faults can be examined over time.
However, when dealing with third-party sensors, we cannot guarantee these assumptions: sensors may start and stop reporting uncontrollably, and the engineer may not have the opportunity to examine the sensors and see how they may fail before having to use their readings.

In contrast to these previous works, the present paper presents a method for sensor fault detection where no model of faulty behavior is available.
As we shall see, the construction of the particle filter gives us the data to perform a particular hypothesis test for sensor fault detection.
This test forms the basis of our fault detection method.
This particular form of hypothesis test is nonparametric, as opposed to \cite{tadic_particle_2014}'s parametric hypothesis test, and only needs a model of non-faulty sensor behavior.
This method can be used in real-time to detect faults in third-party sensors that may not have a fault model, and may sporadically or unpredictably report.

Section \ref{sec:probstatement} gives a mathematical statement of the fault detection problem, and Section \ref{sec:fault_detection} discusses the extension of filtering and state estimation to include our fault detection formulation.
Section \ref{sec:empirical_proportionality} introduces the constraint of a lack of a fault model, and outlines where, mathematically, this prevents us from both state estimation and fault detection.
Section \ref{sec:pf_implementation} introduces the particle filter, and Section \ref{sec:hypothesis} discusses how its structure allows us to naturally overcome the lack-of-fault-model problem. 
Section \ref{sec:case_study} describes an example application, where faulty speed measurements from third-party global navigational satellite system (GNSS) devices are detected in a particle filter for estimating vehicle density on a freeway.

\section{Problem Statement}
\label{sec:probstatement}

Let $x_k \in \mathbb{R}^N$ denote the state vector of our system at time $k$, and $y_k \in \mathbb{R}^{M_k}$ denote the measurement vector at time $k$ (note that $y_k$ may be of different size at different times $k$).
The state and observation vectors evolve over time through discrete-time stochastic state and output equations, denoted $\mathcal{F}_\theta(\cdot)$ and $\mathcal{G}_\theta(\cdot)$ respectively,
\begin{align}
\label{eq:fAndgDeterm}
\begin{split}
  x_k &= \mathcal{F}_\theta \left(x_{k-1} \right) \\
  y_k &= \mathcal{G}_\theta \left(x_k \right),
\end{split}
\end{align}
with $\theta$ a parameter vector describing the randomness or process/measurement noise of $\mathcal{F}$ and $\mathcal{G}$. 
An alternative probabilistic notation may rewrite \eqref{eq:fAndgDeterm} as
\begin{subequations}
\label{eq:xAndYAsRVs}
  \begin{align}
    X_k | \left(X_{k-1} = x_{k-1} \right) 
	    &\sim f\left(x_k | x_{k-1}, \theta \right) \label{eq:fGeneral} \\
    Y_k | \left(X_k = x_k \right) 
	    &\sim g\left(y_k | x_k, \theta \right), \label{eq:gGeneral}
  \end{align}
\end{subequations}
where $X_k$ $(Y_k)$ denotes a random vector and $x_k$ $(y_k)$ the value of a particular realization.
The functions $f( \cdot )$ and $g( \cdot )$ are the probability density functions (PDFs) induced by $\mathcal{F}_\theta( \cdot)$ and $\mathcal{G}_\theta(\cdot)$, respectively. The initial condition of the system, $x_0$, is assumed fixed or distributed with some known density $p(x_0 | \theta )$. More precisely, $f(x_k | x_{k-1}, \theta)$ is a Markov transition kernel with a distribution on the random vector $X_k | (X_{k-1}=x_{k-1})$, and $g(y_k | x_k, \theta)$ is a typical observation likelihood.
To reduce notational clutter, we omit the symbol $\theta$ from now on.

In \eqref{eq:xAndYAsRVs}, the observation PDF is a joint PDF over all elements of the observation vector $Y_k$.
This is the most general formulation of this PDF, and allows for all elements of $Y_k$ to be statistically dependent.
This would be appropriate, if, for example, the entire vector $Y_k$ was reported by a single sensor, and the noise in measuring one element of $Y_k$ was correlated with the noise in measuring the other elements.

In our setting, however, we are dealing with many different sensors and reporting devices.
Say that $Y_k$ contains measurements from $m_k$ different sensors (like $M_k$ (the length of $Y_k$), $m_k$ may be different at different timesteps $k$), and let $j \in \{1, \dots, m_k\}$ index the sensors that report at time $k$ (we will not have $m_k = M_k$ if any sensor reports more than one measurement).
Say that $Y_{k_j}$ is the random vector that contains the element(s) of $Y_k$ from sensor $j$ at time $k$.
Now, if we assume that individual sensors $j$ have independent measurement noises, we may rewrite \eqref{eq:gGeneral} as
\begin{align}
  Y_k | \left(X_k = x_k \right) \sim g\left(y_k | x_k \right) = \prod_{j=1}^{m_k} g_{k_j} \left( y_{k_j} | x_k \right), \label{eq:perSensorG}
\end{align}
where $y_{k_j}$ is the measurement(s) received from sensor $j$ at time $k$ (i.e., a realization of the random vector $Y_{k_j}$), and $g_{k_j}(\cdot)$ is the PDF for $Y_{k_j}$.
Equation \eqref{eq:perSensorG} is an assumption on the form of the joint sensing PDF and is a common assumption in multisensor filtering and sensor fusion (e.g., \cite[(6)]{chen_pf_sensorfusion_04}, \cite[(9)]{chamberland_decentralized_detection_03}, \cite[(35.3)]{durrantwhite_multisensor_16}, \cite[Section 2.2]{mihaylova2007PF})).
We have factored $g(\cdot)$ into its component per-sensor PDFs.

As written in \eqref{eq:perSensorG}, $g_{k_j}(\cdot)$ is in a general form, with no explicit model of faulty vs. non-faulty measurements.
For our purposes, ``faulty'' means any measurement that does not provide information about $x_k$ and that we would like to discard.
We now describe how we handle faulty sensors by explicitly indicating whether sensor $j$ at time $k$ is reporting in either a valid (i.e., as-intended) mode, or a faulty mode.
To this end, we introduce a Bernoulli random variable $Z_{k_j}$, which takes value 1 in the event that a measurement from sensor $j$ is not faulty, and 0 in the event that a measurement is faulty.
Written as a probability mass function (PMF),
\begin{align}
\begin{split}
  p(z_{k_j} | x_k) =& \left( \phi_{k_j}(x_k) \right)^{z_{k_j}}
  \left( 1 - \phi_{k_j}(x_k) \right)^{\left(1- z_{k_j}\right)} \\
   &\quad\textnormal{for } z_{k_j} \in \{0,1\},
   \label{eq:z_pmf}
\end{split}
\end{align}
where $\phi_{k_j}$ is a weight equal to the prior probability that sensor $j$ reports a measurement in a valid, as-intended manner.
It is a \emph{prior} probability, in that it is the probability that any given measurement is non-faulty before we actually see the measurement.
It can be based on prior knowledge, or estimated in real time.

The parameter $\phi_{k_j}$ being a function of $x_k$ allows for the possibility that the $Z_{k_j}$ are dependent on $X_k$; that is, that certain areas of the state space for $x_k$ could lead to a greater likelihood of faulty measurements than others.
This is the most general formulation, and we may simplify the problem by assuming independence of $Z_{k_j}$ from $X_k$ later.

Now, we can condition $g_{k_j}(\cdot)$ from \eqref{eq:perSensorG} on $Z_{k_j}$,
\begin{equation}
  g_{k_j}(y_{k_j} | z_{k_j}, x_k) \triangleq g_{k_j}^{z_{k_j}}(y_{k_j} | x_k),
  \label{eq:y_given_z}
\end{equation}
where $g_{k_j}^1(\cdot)$ is the PDF for the valid sensor behavior, and $g_{k_j}^0(\cdot)$ is the PDF for faulty measurements.
In practice, we should have a model for $g_{k_j}^1(\cdot)$, the expected behavior of the sensor, but $g_{k_j}^0(\cdot)$ may be unknown if we cannot predict every way in which the sensor may fail
 and/or the probabilities of each failure mode.

\section{Extending filtering to include $Z_k$}
\label{sec:fault_detection}
In this Section, we reintroduce recursive, or Bayesian, filtering in a form that includes the variable $Z_k$ we have just defined.
At this point we do not consider the unavailability of $g_{k_j}^0(\cdot)$: that will be considered beginning in Section \ref{sec:empirical_proportionality}.
\subsection{Reformulation of Bayesian Filtering}
In this formulation, detecting a fault for a particular sensor $j$ at time $k$ means determining whether $z_{k_j}$ is equal to 0 or 1.
Of course, $z_{k_j}$ is not directly observable, and its value must be estimated alongside $x_k$ from the observed $y_{k_j}$.

This can be written as a slight extension of traditional recursive Bayesian filtering.
This extension is very similar to filtering for a hybrid system as used in, e.g., \cite{de_freitas_rao-blackwellised_2002}.
Let $Z_k$ represent the collection of random variables $Z_{k_j}$ at time $k$, and let $y_K = \{y_k, y_{k-1}, \dots, y_1\}$, i.e. the set of all measurements up to time $k$. Assume that at time $k$ we have an estimate of the PDF $p(z_{k-1}, x_{k-1} | y_{K-1})$.
Then, we can predict $Z_k$ and $X_k$ before observing measurements $Y_k$,
\begin{align}
  p(z_k, x_k | y_{K-1}) =& p( z_k | x_k) \, p( x_k | y_{K-1}) \nonumber \\
%  \begin{split}
  =& p( z_k | x_k) \bigg( \sum_{z_{k-1}} \int \! p(z_{k-1}, x_{k-1} | y_{K-1}) \nonumber \\
  &\hphantom{p( z_k | x_k) \int } \times f(x_k | x_{k-1}) dx_{k-1} \bigg), \label{eq:predict_with_z}
%  \end{split}
\end{align}
where we have used the Markov property of our system to have $p(z_k | x_k) = p(z_k | x_k, y_{K-1})$, and marginalized out the variables $X_{k-1}$ and $Z_{k-1}$.
Often, when $f(\cdot)$ is nonlinear, computing the integral in closed form in \eqref{eq:predict_with_z} is difficult, and an approximation is used.
One such approximation, the particle filter approximation, is discussed in Section \ref{sec:pf_implementation}.

After computing \eqref{eq:predict_with_z}, we update our predictions once measurements $Y_k$ have been received,
\begin{align}
  p(z_k, x_k | y_K)   &= \frac{p( z_k, x_k, y_k | y_{K-1})}{p( y_k | y_{K-1})} \nonumber \\
  &=  \frac{p( z_k, x_k | y_{K-1} ) p( y_k | x_k, z_k)}{
    p( y_k | y_{K-1} )} \label{eq:filter_with_z}
\end{align}
where we again use the Markov property in writing $p( y_k | x_k, z_k) = p( y_k | x_k, z_k, y_{K-1})$.
Equations \eqref{eq:predict_with_z} and \eqref{eq:filter_with_z} are the prediction and filtering steps, respectively, of recursive Bayesian filtering \cite[Section 2.2]{doucet2011tutorial}, with a slight modification in that $Z_k$ is added.
We will outline how these equations and PDFs are different from the standard formulation.

The function $p( z_k | x_k )$ is the prior PMF of fault/non-fault sensor behavior (prior, in that it is the PMF of $Z_k$ before the measurement vector $Y_k$ is seen).
The PDF $p( y_k | x_k, z_k )$ is the joint likelihood of this measurement vector $Y_k$ from our unobserved system variables $X_k$ and $Z_k$.
Since, in \eqref{eq:perSensorG}, we assumed that the sensors $j$ at time $k$ had independent measurement noises, and could factor the observation model across sensors, we will do the same thing with these two functions,
\vspace{-5pt}
\label{eq:per_sensor_y_and_z}
\begin{align*}
  p( z_k | x_k ) &= \prod_{j=1}^{m_k} p(z_{k_j} | x_k) \displaybreak[0]\\
  p( y_k | x_k, z_k ) &= \prod_{j=1}^{m_k} g_{k_j}^{z_{k_j}} (y_{k_j} | x_k),
\end{align*}
with $p(z_{k_j} | x_k)$ and $g_{k_j}^{z_{k_j}} (y_{k_j} | x_k)$ given by \eqref{eq:z_pmf} and \eqref{eq:y_given_z}, respectively.

The marginal likelihood,
\begin{align}
%\begin{split}
  p(y_k | y_{K-1}) =& \sum_{z_{k-1}} \sum_{z_k} \iint p( z_{k-1}, x_{k-1} | y_{K-1}) f( x_k | x_{k-1}) \nonumber \\
  &\times p( z_k | x_k) p( y_k | x_k, z_k) dx_{k-1} dx_k, \label{eq:marginal_likelihood}
%  &\times dx_{k-1} dx_k \nonumber
%\end{split} 
\end{align}
plays the role of a normalizing constant.

\subsection{Relevant conditional and marginal PDFs}
The PDFs \eqref{eq:predict_with_z} and \eqref{eq:filter_with_z} are the joint PDFs of the state $X_k$ and sensor fault probabilities $Z_k$ conditioned on measurements.
These joint PDFs, however, are high dimensional, which presents two important drawbacks.

First, the PDFs can become very large. If the size of an estimate of $p( x_k | y_k)$ is proportional to $N$ (the dimensionality of $X_k$), the size of an estimate of the joint PDF $p( z_k, x_k | y_K)$ will be proportional to $N\cdot 2^{M_k}$ (recall $M_k$ is the length of $Y_k$), since a different estimate of $X_k$ will exist for every combination of faulty/not faulty of all sensors (cf. \cite{wei_adaptive_2009}).
For all but trivial problems, this may be infeasible.

Second, they are difficult to interpret: these large joint probabilities do not immediately answer the questions ``what is the probability that sensor $j$ at time $k$ is faulty'' or ``what is the best estimate of the state $X_k$ after accounting for potential faults.''

Both concerns are answered with marginal or conditional PDFs for $Z_{k_j}$ and $X_k$, respectively.
We discuss how to compute these marginal and conditional PDFs in this Section.

One of these values, the posterior probability that a particular sensor $j$ at time $k$ is faulty for a particular value of $X_k$ has the PMF
\vspace{-5pt}
\begin{align}
  &\begin{aligned}
    p( z_{k_j} | x_k, y_K) &= p( z_{k_j} | x_k, y_{k_j}) = \frac{p( z_{k_j} | x_k) p(y_{k_j} | x_k, z_{k_j})}{ p( y_{k_j} | x_k)} \nonumber \\
%  &= \frac{p( z_{k_j}, x_k | y_{k-1}, \theta) g_{k_j}^{z_k_j} ( y_{k_j} | x_k, \theta )}{
%    p( y_k | y_{k-1}, \theta) p( x_k | y_k, \theta )} \nonumber \\
    &= \frac{p( z_{k_j} | x_k) g_{k_j}^{z_{k_j}} ( y_{k_j} | x_k )}{
     p( y_{k_j} | x_k)}
     \end{aligned} \nonumber \displaybreak[0]\\
  &\begin{aligned}
  = &\frac{\left( \phi_{k_j}(x_k) \right)^{z_{k_j}}
  \left( 1 - \phi_{k_j}(x_k) \right)^{\left(1- z_{k_j}\right)}
  g_{k_j}^{z_{k_j}} ( y_{k_j} | x_k )}{
    p( y_{k_j} | x_k)} \\
   &\qquad\textnormal{for } z_{k_j} \in \{0,1\},
  \end{aligned}    \label{eq:posterior_z_given_x}
%  &=  \begin{cases}
%        \begin{aligned}
%          &\phi_{k_j}(x_k) \\
%          &\times g_{k_j}^1 (y_{k_j} | x_k, \theta) 
%        \end{aligned} & \textnormal{for } z_{k_j} = 1 \\
%        \begin{aligned}
%          &\left(1 - \phi_{k_j}(x_k) \right) \\
%          &\times g_{k_j}^0 (y_{k_j} | x_k, \theta)
%        \end{aligned} & \textnormal{for } z_{k_j} = 0.
%  \end{cases}
\end{align}
where in the first equality we have again used the assumption that the sensors $j$ at time $k$ have independent measurement noises and the Markov property of $f(\cdot)$, and the final two equalities use \eqref{eq:z_pmf} and \eqref{eq:y_given_z}.

From \eqref{eq:posterior_z_given_x}, we can see that determining the probability that a sensor is faulty is itself a Bayesian inference problem, with $p( z_{k_j} | x_k)$ being a prior distribution and $g_{k_j}^{z_{k_j}} ( y_{k_j} | x_k )$ a likelihood.
The marginal likelihood in the denominator,
\begin{align}
% \begin{split}
p(y_{k_j} | x_{k}) =& \sum_{z_{k_j}} p( z_{k_j} | x_{k}) p( y_{k_j} | x_k, z_{k_j}) \nonumber \displaybreak[0]\\
=& \sum_{z_{k_j}} p( z_{k_j} | x_{k})
 g_{k_j}^{z_{k_j}}(y_{k_j} | x_k),
\label{eq:z_marginal_likelihood}
% \end{split}
\end{align}
again acts as a normalizing constant.

The PMF \eqref{eq:posterior_z_given_x} is a conditional probability on $X_k$: it gives the probability that sensor $j$ at time $k$ is faulty for a \emph{particular} value of $X_k$.
The marginal PMF
$p(z_{k_j} | y_K)$, which averages the fault probability over all possible values of $X_k$; and PDF $p(x_k | y_K)$, which weights our state estimate by the estimated probability of fault for each sensor; are easy to calculate from \eqref{eq:filter_with_z},
\begin{subequations} \label{eq:marginals}
  \begin{align}
    p( z_{k_j} | y_K) &= \int \sum_{z_k \setminus z_{k_j}} p( z_{k}, x_k | y_K) dx_k \label{eq:posterior_z} \displaybreak[0]\\
    p( x_k | y_K) &= \sum_{z_k} p(z_k, x_k | y_K). \label{eq:integrate_out_z} 
  \end{align}
\end{subequations}

\section{Problem statement without a model of faults}
\label{sec:empirical_proportionality}
Recall from Section \ref{sec:probstatement} that we may only have a model of $g_{k_j}^{z_{k_j}} ( y_{k_j} | x_k )$ of the case $Z_{k_j}=1$ (the measurement PDF of a sensor operating correctly), if we cannot predict every way a measurement might be faulty.
This Section discusses how not having a model of faulty behavior stymies Section \ref{sec:fault_detection}'s straightforward attempts at filtering.

Note that $g_{k_j}^0(\cdot)$ appears in all of the PDFs of interest in Section \ref{sec:fault_detection}: explicitly as in \eqref{eq:posterior_z_given_x} or within the summations over $z_k$ that appear in the marginal likelihoods \eqref{eq:marginal_likelihood} and \eqref{eq:z_marginal_likelihood}.
Not having a model of $g_{k_j}^0(\cdot)$ means that these equations cannot actually be computed.
In the case of $p(z_{k_j} | x_k, y_{k_j})$, it is simple enough to restrict ourselves to $Z_{k_j}=1$ and compute \eqref{eq:posterior_z_given_x} in that special case,
\begin{align}
  P( Z_{k_j} \!=\! 1 | X_k \!=\! x_k, \! Y_{k_j} \!=\! y_{k_j}) \! = \!
   \frac{\phi_{k_j}(x_k)
  g_{k_j}^1 ( y_{k_j} | x_k )
  }{p(y_{k_j} | x_k)}.
%     \label{eq:prob_of_nonfault}
\nonumber
\end{align}

However, the denominator $p(y_{k_j} | x_k)$, the marginal likelihood and normalizing constant, still requires $g_{k_j}^0(\cdot)$ to compute it from \eqref{eq:z_marginal_likelihood}.
With what we have so far, the best we can do is to emphasize that the marginal likelihood is a constant since it does not vary with $Z_{k_j}$,
\begin{align}
\begin{split}
  P( Z_{k_j} = 1 | X_k = x_k, Y_{k_j} &=  y_{k_j}) \\
   =& \; C_{k_j} \phi_{k_j}(x_k)
  g_{k_j}^1 ( y_{k_j} | x_k ), \label{eq:prob_of_nonfault_with_const}
%   \nonumber
%   \begin{split}
%     &P( Z_{k_j} \!=\! 1 | Y_{k_j} \!=\! y_{k_j}) \! = 
%     \begin{aligned}[t]
%     C \int& \phi_{k_j}(x_k) g_{k_j}^1 ( y_{k_j} | x_k ) \\
%     &\times p(x_k | y_{k-1}) dx_k
%     \end{aligned}
%   \end{split}
\end{split}
\end{align}
for some scalar constant $C_{k_j}$.

Parallel analyses are possible for the joint PDF $p(z_k, x_k | y_K)$ and the marginals in \eqref{eq:marginals}, with the unsolvable marginal likelihood in these cases being $p(y_k | y_{K-1})$ in \eqref{eq:marginal_likelihood}.

In \eqref{eq:prob_of_nonfault_with_const}, we have replaced the normalizing denominator containing $g_{k_j}^0(\cdot)$ with a constant that must be estimated.
This can be thought of as a general, and difficult, parameter estimation or adaptive control problem. The best parameter estimation method for $C_{k_j}$ will vary on a case-by-case basis with the particular system considered and available data.

As it turns out, though, the inability to calculate normalizing denominators is a particularly common issue in Bayesian filtering: marginal likelihoods like \eqref{eq:marginal_likelihood} and \eqref{eq:z_marginal_likelihood}
often cannot be computed when the PDF $f( x_k | x_{k-1})$ is not available in closed form or not easily integrable.
In these situations, the \emph{particle filtering} algorithm is appropriate, as it can approximate the prediction and filtering steps without computing $p( y_k | y_{K-1})$.
The next Section introduces the particle filtering algorithm, and Section \ref{sec:hypothesis} discusses how to extend it to deal with the inability to directly compute \eqref{eq:marginal_likelihood} or \eqref{eq:z_marginal_likelihood} due to a lack a model of faults.

\section{Particle Filter Implementation}
\label{sec:pf_implementation}
The methods discussed so far make use of many PDFs (and integrals thereof).
These operations may be performed in closed form when the integrals are relatively simple, but this is often not the case when the system \eqref{eq:fAndgDeterm} is nonlinear.

For these situations, state estimation is often performed using Monte Carlo methods, with perhaps the most widely-used method being the particle filter \cite{doucet2011tutorial}.
A particle filter may be used when no closed-form model for $f(\cdot)$ exists (but the PDF may be sampled from by, e.g., running a stochastic simulation many times), or when numerically computing the integrals in \eqref{eq:predict_with_z} and \eqref{eq:marginal_likelihood} is computationally expensive.

A particle filter is constructed by replacing PDFs for $X_k$ and $Z_k$ with approximate PDFs (denoted with a hat) made up of many discrete samples (also called particles) from the continuous PDF.
We extend a traditional particle filter by including the random variable $Z_k$ from \eqref{eq:predict_with_z} (again, this is similar to previous works, e.g., \cite{de_freitas_rao-blackwellised_2002}).
Starting with an approximate PDF from the previous timestep, $\hat{p}(z_{k-1}, x_{k-1}| y_{K-1})$,
\begin{align}
  &\begin{aligned}[t]
   p(z_k, x_k | y_{K-1}) =
   p( z_k | x_k) \bigg( & \sum_{z_{k-1}} \int \! p(z_{k-1}, x_{k-1} | y_{K-1}) \nonumber \\
  & \times f(x_k | x_{k-1}) dx_{k-1} \bigg) \nonumber
  \end{aligned} \displaybreak[0]\\
  &\approx 
  \begin{aligned}[t]
    p(z_k | x_k) \sum_{p=1}^P \bigg(  \sum_{z_{k-1}} p( z_{k-1}, x_{k-1}^p | y_{K-1} ) 
    \delta_f (x_k^p |x_{k-1}^p  ) \bigg)
  \end{aligned} \nonumber \displaybreak[0]\\
  &= p(z_k | x_k) \hat{p}(x_k | y_{K-1}) \nonumber \displaybreak[0]\\
  &= \hat{p}(z_k, x_k | y_{K-1}), \label{eq:pf_predict}
\end{align}
where $P$ is an integer denoting the total number of particles drawn from $f( \cdot )$, $p \in \{1,\dots,P\}$ indexes individual particles (or atoms of the probability distribution), and $\delta_f \left(x^p_k |x_{k-1}^p \right)$ is a Dirac delta, which places a unit mass on the point $x_k^p|x_{k-1}^p$, which is in turn equal to $\mathcal{F}( x_{k-1}^p)$ for the $p$th particle.
The term $p( z_{k-1}, x_{k-1}^p | y_{K-1})$ refers to the probability for this particle $p$ from our approximate PDF from the prior timestep, $\hat{p}(z_{k-1}, x_{k-1}| y_{K-1})$.
The final two equalities indicate that the empirical PDF $\hat{p} (z_k, x_k|y_{K-1})$ consists of a weighted sum of $P$ points $(z_k | x_k^p, \; x_k^p |x_{k-1}^p)$, with individual weights $p(x_k^p  | y_{K-1})$, where the weights sum to one.
% A straightforward application of the strong law of large numbers shows that as $P \to \infty$, $\hat{p} \left(z_k, x_k | y_{k-1} \right) \to p\left(z_k, x_k | y_{k-1} \right)$ almost surely~\cite{doucet2011tutorial}.

The corresponding particle filter update equation comes from plugging \eqref{eq:pf_predict} into \eqref{eq:filter_with_z},
\begin{align}
  &
  \begin{aligned}
  p(z_k, x_k | y_K) &= \frac{p( z_k, x_k | y_{K-1} ) p( y_k | x_k, z_k)}{
  p( y_k | y_{K-1} )} \nonumber \\
  &\approx \frac{\hat{p}( z_k, x_k | y_{K-1} ) p( y_k | x_k, z_k)}{
  p( y_k | y_{K-1} )}
  \end{aligned} \nonumber \displaybreak[0]\\
  &\begin{aligned}
    = \frac{1}{p(y_k | y_{K-1})} \bigg( &\sum_{p=1}^P p(z_k | x_k^p) p(y_k | x_k^p, z_k^p) \delta_f (x_k^p |x_{k-1}^p  ) \nonumber \\
    &\times \sum_{z_{k-1}} p( z_{k-1}^p, x_{k-1}^p | y_{K-1} ) \bigg)
    \end{aligned} \nonumber \displaybreak[0]\\
  % &= \frac{1}{p(y_k | y_{K-1})} \sum_{p=1}^P p(z_k^p, x_k^p | y_k) \delta_f (x_k^p |x_{k-1}^p  ) \label{eq:pf_filter} \displaybreak[0]\\
  % 
  &= \frac{\sum_{p=1}^P p(z_k^p, x_k^p | y_k) \delta_f (x_k^p |x_{k-1}^p  )}{p(y_k | y_{K-1})} \label{eq:pf_filter} \displaybreak[0]\\
  &= \hat{p}( z_k, x_k | y_K). \nonumber
\end{align}
This posterior empirical PDF $\hat{p}( z_k, x_k | y_K)$ is thus made of the same collection of Dirac deltas as in $\hat{p}( z_k, x_k | y_{K-1})$, but with updated weights to reflect each point's updated probability (i.e., the prior probability times the likelihood).

As mentioned above, use of a particle filter also allows us to avoid having to calculate the marginal likelihood $p(y_k | y_{K-1})$ using \eqref{eq:marginal_likelihood}.
Instead, after $p(z_k^p, x_k^p | y_K)$ is calculated for every particle $p$ in \eqref{eq:pf_filter}, we normalize these probabilities so that they sum to one,
% \begin{align}
%   \hat{p}( z_k, x_k | y_k) = \sum_{p=1}^P \frac{ p(z_k^p, x_k^p | y_k) \delta_f (x_k^p |x_{k-1}^p  )}{\sum_{p=1}^P p(z_k^p, x_k^p | y_k) }. \label{eq:pf_normalization}
% \end{align}
% In other words,
\begin{align}
  p(y_k | y_{K-1}) \approx \sum_{p=1}^P p(z_k^p, x_k^p | y_K). \label{eq:pf_normalization}
\end{align}
Equations \eqref{eq:pf_filter}-\eqref{eq:pf_normalization} make up the update computation that is used in practice.
However, for fault detection we will use a different update step, detailed in the next Section.

For brevity, we omit discussion of the particle filter's post-update resampling step.
See, e.g., \cite{doucet2011tutorial}, for details.

\section{Particle-filter-enabled nonparametric hypothesis testing}
\label{sec:hypothesis}
\subsection{Estimating the PDF of our null hypothesis}
We now discuss the use of particle filter for sensor fault detection without a model of faults, $g_{k_j}^0(\cdot)$.
To do so, we propose a modification described in this Section.
Once we have computed a prediction step \eqref{eq:pf_predict}, instead of the traditional update step \eqref{eq:pf_filter}, we examine individually whether each sensor $j$ at time $k$ can be determined to be faulty.
For every sensor, we can find
\begin{align}
&p( y_{k_j} | Z_{k_j} = 1, y_{K-1}) = \int p( y_{k_j}, x_k | Z_{k_j}=1, y_{K-1}) dx_k \nonumber \displaybreak[0]\\
&= \int p( y_{k_j} | Z_{k_j} = 1, x_k) p( x_k | Z_{k_j} = 1, y_{K-1}) dx_k \nonumber \displaybreak[0]\\
&= \int g^1_{k_j}( y_{k_j} | x_k) \frac{p( x_k, Z_{k_j}=1 | y_{K-1})}{P(Z_{k_j} = 1 | y_{K-1})} dx_k \nonumber \displaybreak[0]\\
% &= \begin{aligned}[t]
%   &\frac{1}{P( Z_{k_j}=1 | y_{K-1})} \nonumber \\
%   &\times \int g^1_{k_j} ( y_{k_j} | x_k) P( Z_{k_j} = 1 | x_k) p( x_k | y_{K-1}) dx_k
%   \end{aligned} \displaybreak[0]\\
% 
&= \int g^1_{k_j} ( y_{k_j} | x_k) \frac{P( Z_{k_j} = 1 | x_k) p( x_k | y_{K-1})}{P( Z_{k_j}=1 | y_{K-1})} dx_k \nonumber \displaybreak[0]\\
  % 
%   \begin{split}
% &\approx \begin{aligned}[t]
%   &\frac{1}{P( Z_{k_j}=1 | y_{K-1})} \\
%   &\times \sum_{p=1}^P g^1_{k_j} (y_{k_j} | x_k^p) \phi_{k_j}(x_k^p) p( x_k^p | y_{K-1})
%   \end{aligned} \label{eq:likelihood_on_nonfault}
%   \end{split} \displaybreak[0]\\
% 
&\approx
  \frac{\sum_{p=1}^P g^1_{k_j} (y_{k_j} | x_k^p) \phi_{k_j}(x_k^p) p( x_k^p | y_{K-1})}{P( Z_{k_j}=1 | y_{K-1})} \label{eq:likelihood_on_nonfault} \displaybreak[0]\\
&= \hat{p}(y_{k_j} | Z_{k_j} = 1, y_{K-1}), \nonumber
\end{align}
where, now, with $y_{k_j}$ \emph{not} being conditioned on, $g^1_{k_j} (y_{k_j} | x_k^p)$ means the actual PDF for $Y_{k_j} | (X_k^p = x_k^p, Z_{k_j}=1)$.
Therefore, the PDF $\hat{p}(y_{k_j} | Z_{k_j} = 1, y_{K-1})$ is now a sum of non-delta PDFs; it can be compared to a kernel density estimate \cite[Ch. 6]{scott2015multivariate}.
The term $p( x_k^p | y_{K-1})$ in \eqref{eq:likelihood_on_nonfault} is just the particle $p$'s prior probability from $\hat{p}(z_k, x_k | y_{K-1})$.

The sum over $p( x_k^p | y_{K-1})$ for particles $p$ weighs this PDF by the probabilities of particular values of $x_k$.
Note that, since we have conditioned on $Z_{k_j} = 1$, the constant $C_{k_j}$ from \eqref{eq:prob_of_nonfault_with_const} does not appear.
The denominator $P(Z_{k_j} = 1 | y_{K-1})$ is unknown, and computing it in closed from requires integrals over $x_k$.
However, taking advantage of the particle filter structure, we can approximate this term similarly to \eqref{eq:pf_normalization},
\begin{equation}
P(Z_{k_j}=1 | y_{K-1}) \approx \sum_{p=1}^P \phi_{k_j}(x_k^p) p( x_k^p | y_{K-1}) \label{eq:z_pf_normalization}.
\end{equation}

The empirical PDF in \eqref{eq:likelihood_on_nonfault}-\eqref{eq:z_pf_normalization} is an estimate of $p(y_{k_j} | Z_{k_j}=1, y_{K-1})$, the PDF of sensor $j$ at time $k$'s measurements under non-faulty sensor behavior.
We will use this PDF to evaluate whether specific sensors might be faulty.

\subsection{Performing the hypothesis test}
Plugging a particular value of $y_{k_j}$ into $p(y_{k_j} | Z_{k_j} = 1, y_{K-1})$, assuming $Y_{k_j}$ is a continuous random variable or vector, gives (loosely speaking) $P(Y_{k_j} \in B_\epsilon(y_{k_j}) | Z_{k_j} = 1, y_{K-1})$, where $B_\epsilon(y_{k_j})$ is an $\epsilon$-ball centered at $y_{k_j}$.
Finding the value of our estimate $\hat{p}(y_{k_j} | Z_{k_j} = 1, y_{K-1})$ at an observed $y_{k_j}$, then, gives $\hat{P}(Y_{k_j} \in B_\epsilon(y_{k_j}) | Z_{k_j} = 1, y_{K-1})$.

At this point, we have established the necessary components to conduct a hypothesis test for a null hypothesis $H_0: Z_{k_j} = 1$.
For some $\alpha \in (0,1)$, we can reject $H_0$ and conclude that $Z_{k_j}=0$ (that is, that the sensor is faulty and should be excluded) if
\begin{equation}
%   \hat{p}( y_{k_j} | Z_{k_j} = 1) =
  \hat{P}(Y_{k_j} \in B_\epsilon(y_{k_j}) | Z_{k_j} = 1, y_{K-1}) < \alpha \label{eq:hypothesis_test}
\end{equation}
for the observed value of $y_{k_j}$.
% \footnote{
% Like other typical particle filter estimates, the rate of convergence of the estimator in \eqref{eq:hypothesis_test} to the true value as $P$ increases (and hence the number of particles needed for ``good'' performance), will depend on the dynamics and stochasticity of the particular system considered.}

The $\alpha$ here plays the role of a \emph{significance level} for the hypothesis test, which is the maximum acceptable probability of declaring a sensor faulty when it is actually non-faulty.
Here, the test in \eqref{eq:hypothesis_test} is a \emph{nonparametric} hypothesis test, as we use a nonparametric model of $p(y_{k_j} | Z_{k_j} = 1, y_{K-1})$ as given by \eqref{eq:likelihood_on_nonfault}-\eqref{eq:z_pf_normalization}\footnote{
Note that, strictly speaking, \eqref{eq:hypothesis_test}'s hypothesis test is not tied to a particle filter as we have developed it in this paper.
The particle filter, though, gives us a generally-applicable nonparametric hypothesis test.
If we have a system where we know a parametric filter (such as a Kalman or Extended Kalman Filter) is appropriate (i.e., if we have a closed-form, easily-integrable $f(\cdot)$), we could derive \eqref{eq:likelihood_on_nonfault} and \eqref{eq:hypothesis_test} for the alternative filter's parametric PDFs.
}.

Once we have rejected sensors that meet \eqref{eq:hypothesis_test}, the remaining sensors (those for which we could not reject $H_0$) can be concluded to be non-faulty and their measurements can be used in a particle filter update equation that takes $Z_{k_j}=1$ for these remaining sensors:
\begin{align}
  \hat{p}(x_k | y_K) &= \nonumber
%   \begin{aligned}[t]
   \frac{1}{p(y_k | y_{K-1})} \\
  \times \sum_{p=1}^P & \bigg( \prod_{k_j \in \mathcal{H}_0} g^1_{k_j} (y_{k_j} | x_k^p) \bigg)
  \delta_f (x_k^p |x_{k-1}^p  )
    p( x_{k-1}^p | y_{K-1} ) \nonumber \displaybreak[0]\\
    % &= \frac{1}{p(y_k | y_{K-1})} \sum_{p=1}^P p(x_k^p | y_K) \delta_f (x_k^p |x_{k-1}^p  ), \label{eq:new_pf_update}
    %
    &= \frac{\sum_{p=1}^P p(x_k^p | y_K) \delta_f (x_k^p |x_{k-1}^p  )}{p(y_k | y_{K-1})}, \label{eq:new_pf_update}
\end{align}
where $\mathcal{H}_0$ is the set of sensors that did not meet \eqref{eq:hypothesis_test}, and the denominator $p(y_k | y_{K-1})$ is estimated via \eqref{eq:pf_normalization}.

A higher value of $\alpha$ leads to more aggressive rejection of measurements.
Also note that in general we do \emph{not} have $\hat{P}(Y_{k_j} \in B_\epsilon(y_{k_j}) | Z_{k_j} = 0, y_{K-1}) = 1 - \hat{P}(Y_{k_j} \in B_\epsilon(y_{k_j}) | Z_{k_j} = 1, y_{K-1})$.
The RHS, the true probability that a sensor $j$ at time $k$ is taken to be non-faulty when it is actually faulty, needs knowledge or estimation of $g_{k_j}^0(\cdot)$.

\subsection{Algorithm}
Our particle filter method with real-time fault detection without fault models can be summarized as:
\begin{enumerate}
  \item Perform a prediction step as normal, using \eqref{eq:pf_predict}.
  \item For each sensor $j$ at time $k$, calculate $\hat{p}(y_{k_j}~|~Z_{k_j}~=~1,~y_{K-1})$ using \eqref{eq:likelihood_on_nonfault}.
  \item For each sensor, determine whether to reject it as faulty using \eqref{eq:hypothesis_test} for the observed $y_{k_j}$.
  \item Perform an update step with the non-rejected measurements using \eqref{eq:new_pf_update} (and, if desired, a resampling step).
  \item Advance in time, $k \leftarrow k+1$, return to step 1, repeat.
\end{enumerate}

In summary, repurposing of the particles we already have when we use a particle filter allows us to perform hypothesis tests, and detect faults of unknown form, for free.

\section{A case study: GNSS sensor fusion for highway traffic state estimation}
\label{sec:case_study}
Our case study concerns a traffic control system that takes possibly-faulty third-party data from connected vehicles' global navigational satellite system (GNSS) devices to estimate the state of a road network.
Accurate knowledge of traffic systems' operations is needed for reactive traffic control, and
gaining this knowledge from third-party data is a goal of modern intelligent transportation system ``smart city'' applications \cite{work2009trafficmodel, kurvar15}.

\begin{figure*}[ptb]
  \centering
  \subfloat[][True density state (veh/m)]{
    \includegraphics[width=.3\textwidth, trim=0 0 2.5pc 1.5pc,clip]{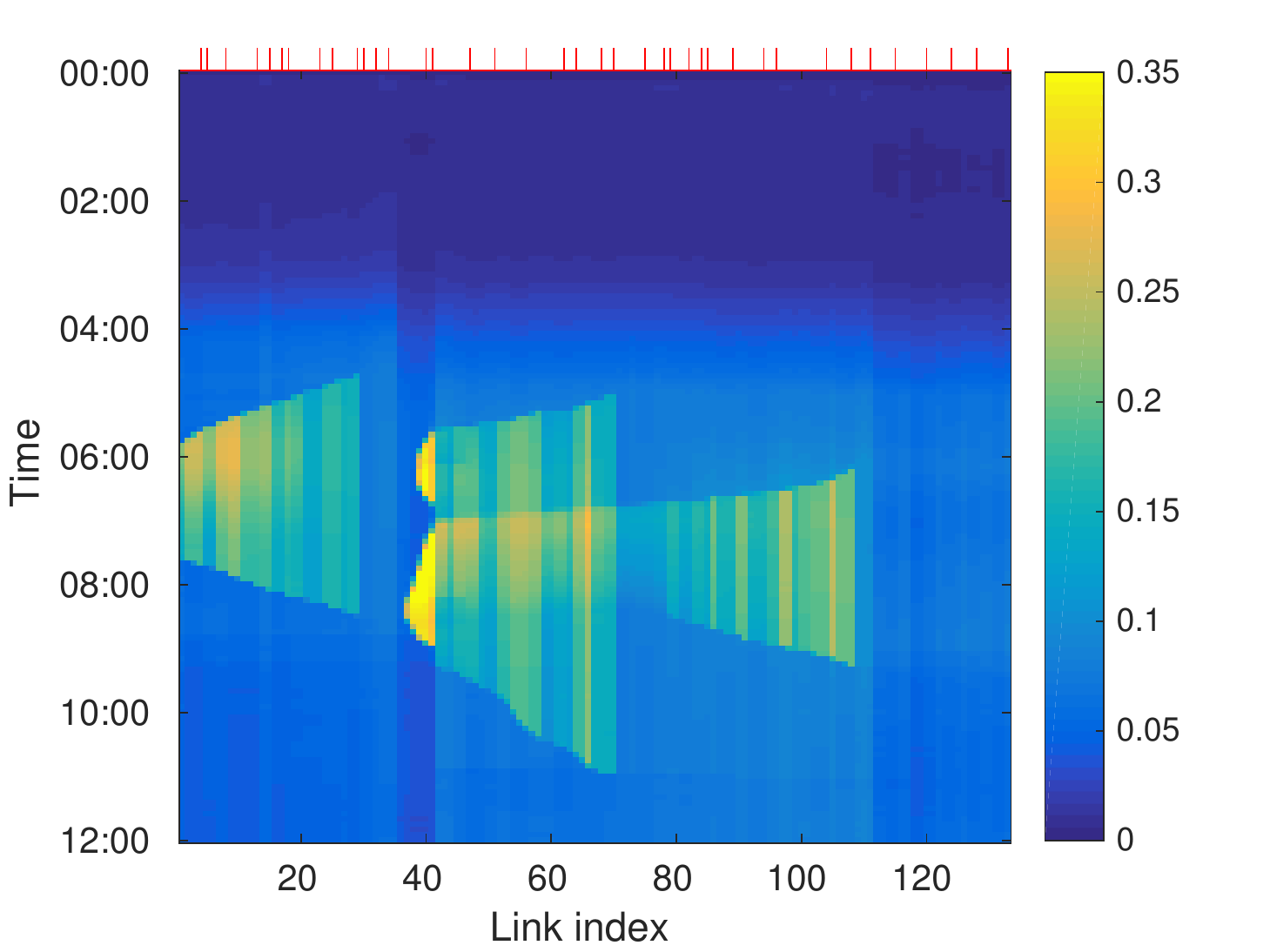}
    \label{fig:c_point_001}
  }
  \subfloat[][Speed measurements (m/s)]{
    \includegraphics[width=.3\textwidth, trim=0 0 2.5pc 1.5pc,clip]{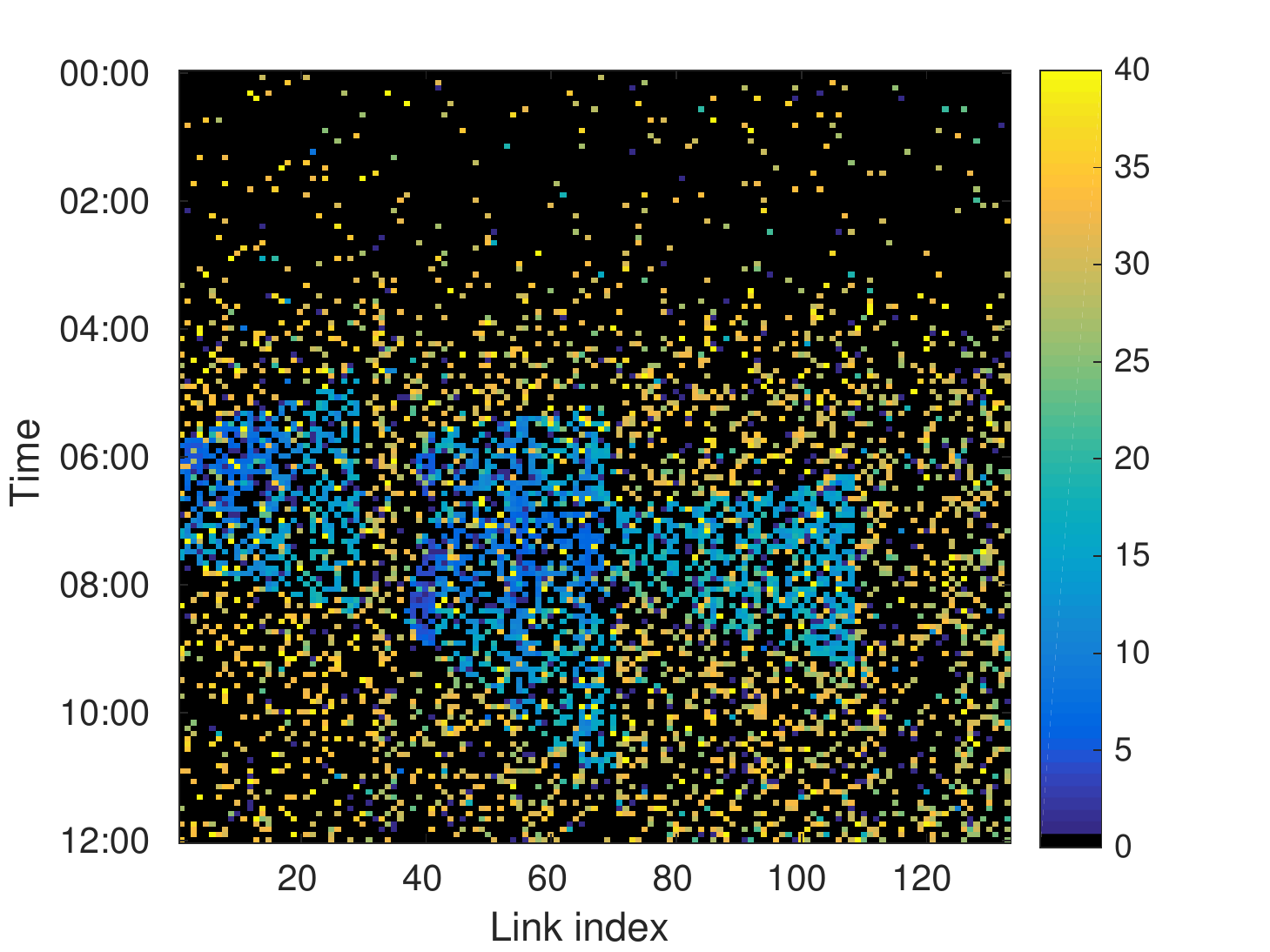}
    \label{fig:c_point_01}
    }
  \subfloat[][Non-faulty speed subset (m/s)]{
    \includegraphics[width=.3\textwidth, trim=0 0 2.5pc 1.5pc,clip]{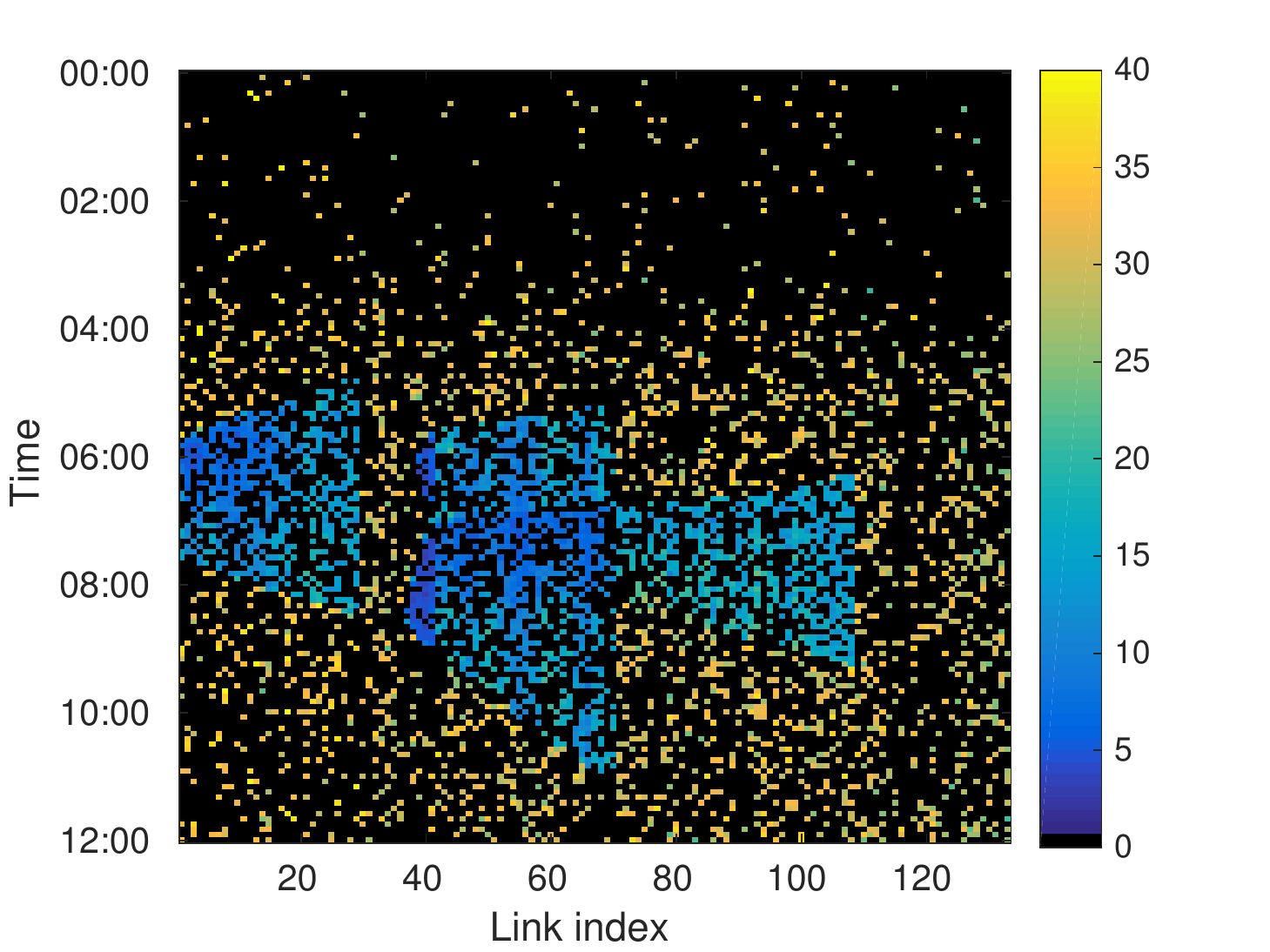}
    \label{fig:c_point_1}
    }
  \caption{Simulated true density state trajectory (a), speed measurements (b), and non-faulty subset of speed measurements (c) used in simulation. Traffic moves to the right, and the time period considered is midnight to noon (as marked on the vertical axis). In (a), the links instrumented with loop detectors that noisily measure density are marked with red ticks. At peak morning demand, bottlenecks near links 30, 70, and 110 lead to traffic jams that propagate upstream (i.e., they extend to the left as time advances), leading to increased density and lower speed. The jams later dissipate as demand falls. \vspace{-12pt}}
  \label{fig:contours}
\end{figure*}

\subsection{System model equations for our case study}
In \cite[Section IV-B]{wright2016fusion}, we presented results obtained using a particle filter that used the techniques detailed in Section \ref{sec:pf_implementation}.
% We only briefly mentioned the use of the fault detection techniques in \cite{wright2016fusion}; this section discusses them in detail.
%GNSS data and flow readings from embedded-in-the-road inductive loops.
% We estimated freeway traffic density using fused third- and first-party data.
We demonstrated the particle filter's performance on a 19-mile portion of I-210 West in southern California.
As our system model $f(\cdot)$, we make use of the macroscopic Cell Transmission Model (CTM) \cite{daganzo94}, which approximates traffic as compressible fluid flows.
This type of model can capture important nonlinear emergent features in traffic flows like traffic jams and congestion waves.

In the CTM, the freeway is discretized into a sequence of finite-volume cells, also called links.
The state vector $x_k$ is the vector of link densities $\rho_{\ell,k}$.
The state update equation for link $\ell$ is
\vspace{-5pt}
\begin{equation}
  \rho_{\ell,k+1} = \rho_{\ell,k} + \frac{1}{L_\ell} (q_{\ell-1,k} - q_{\ell,k} + r_{\ell,k} - s_{\ell,k}), \label{eq:ctm}
\end{equation}
where $L_\ell$ is the length of link $\ell$, $q_{\ell,k}$ denotes the vehicle flow leaving link $\ell$ to link $\ell+1$ at time $k$, $r_{\ell,k}$ is the flow entering link $\ell$ from an onramp (if any) at time $k$, and $s_{\ell,k}$ is the flow leaving link $\ell$ to an offramp (if any) at time $k$.
When there is no onramp entering link $\ell+1$, the inter-link flows in \eqref{eq:ctm} are given by
\begin{align}
\begin{split}
q_{\ell,k} = \min(& v_{f,\ell} \cdot \rho_{\ell,k} \cdot L_\ell, \, Q_{max,\ell},\\
 &w_{\ell+1} \cdot L_{\ell+1} \cdot (\rho_{J,\ell+1} - \rho_{\ell+1,k})), \label{eq:ctmq}
\end{split}
\end{align}
where $v_{f,\ell}$ is the freeflow speed of link $\ell$, $Q_{max,\ell}$ is the capacity, or maximum possible flow over a time period, of link $\ell$, $w_{\ell+1}$ is the speed at which congestion waves propagate upstream in link $\ell+1$, and $\rho_{J,\ell+1}$ is the jam density, or maximum possible density, of link $\ell+1$.
The third term in the $\min(\cdot)$ function in \eqref{eq:ctmq} lets the downstream link $\ell+1$ refuse to accept flow from link $\ell$ if $\ell+1$ is too full.

When there is an onramp entering link $\ell+1$, its available supply (the third argument to \eqref{eq:ctmq}'s $\min(\cdot)$ function) is distributed among link $\ell$ and the onramp according to the junction model of \cite{muralidharan2009lnctm}.
The ramp flows themselves, $s_{\ell,k}$ and $r_{\ell,k}$ in \eqref{eq:ctm}, are random variables.
See \cite{wright2016fusion} for full implementation details of these last two points.

A common type of first-party sensor for freeway traffic are inductive loop detectors buried in the pavement.
These detectors can noisily measure density.
A third-party source of data are vehicle-carried GNSS devices that report the speed of individual vehicles.
In the CTM, the speed of traffic in link $\ell$ at time $k$ is $v_{\ell,k} = L_l \cdot \rho_{\ell,k} / q_{\ell,k}$.
A high vehicle density leads to congestion, and hence low speeds.
We can use speed measurements to estimate density using this relationship in a Rao-Blackwellized particle filter, as discussed in \cite{wright2016fusion}.

To test our fault detection method, we simulated a realization of our freeway model, with randomness introduced by the random onramp, offramp, and upstream boundary flows.
In addition to noisy density measurements from 41 loop detectors, we simulated GNSS speed measurements with a simulated penetration rate of 2\%.
To generate the faulty third-party measurements, we used $\phi_{k_j}(x_k) = 0.7$ for all $j$, $k$, and $x_k$, i.e., each speed measurement had a 30\% probability of being faulty.
We used two fault models: a faulty measurement had a 1/3 probability of reporting zero (i.e., a stopped car misreporting its location), and a 2/3 probability of drawing from a Gaussian distribution with mean 30 m/s and standard deviation 10 m/s.
The non-fault model for velocity measurements, $g_{k_j}^1(\cdot)$, was Gaussian with a mean of the true link velocity and standard deviation of 10\% of the mean (similar to \cite{work2009trafficmodel}).
Fig. \ref{fig:contours} shows the true state and velocity measurements used.
Note that the particle filter state estimator/fault detector has no information of which measurements might be faulty or what faults may look like.

\begin{table}
\caption{Simulated GNSS measurement fault detection results}
\label{tab:results}
\begin{tabular}{l r r r r}
  \toprule
  & \multicolumn{1}{c}{$\alpha = 0.001$} & \multicolumn{1}{c}{$\alpha = 0.01$} &  \multicolumn{1}{c}{$\alpha = 0.1$} & No Faults \\ \cmidrule(r){2-5}
  True Positives & 1092 & 1340 & 1994 & - \\
  False Positives & 88 & 95 & 2868 & - \\
  True Negatives & 4506 & 4499 & 1726 & - \\
  False Negatives & 914 & 666 & 12 & - \\
  Labeling Error & 15.18\% & 11.53\% & 43.64\% & - \\ \midrule
  Density MAPE & 3.81\% & 3.51\% & 4.63\% & 3.43\% \\
  \bottomrule
\end{tabular}
\vspace{3pt}

{\footnotesize In these results, an incorrect $\phi_{k_j} = 0.3$ is used (see text). ``Positives'' refer to sensors for which we accepted $H_1$, i.e., sensors that our fault detection hypothesis test \eqref{eq:hypothesis_test} concluded were faulty. ``True'' and ``False'' refer to correct and incorrect decisions, respectively, of whether a sensor is faulty. \\MAPE = mean absolute percentage error.}
\vspace{-12pt}
\end{table}

\subsection{Results}
Table \ref{tab:results} presents the results of estimation with our fault detection method for several levels of $\alpha$, as well as a ``best case'' state estimator where the faulty measurements are not present.
In these fault detection tests with our previously-generated measurements, we used $\phi_{k_j}$ = 0.3 in the estimation algorithm.
This $\phi_{k_j}$ value was chosen as an arbitrary example of an incorrect $\phi_{k_j}$ value, since the true $\phi_{k_j}$ for this example problem was 0.7 (as set in the problem specification).
By running our algorithm with an incorrect $\phi_{k_j}$ value, we demonstrate that the hypothesis test \eqref{eq:hypothesis_test} is robust to very large uncertainty in this prior probability in this application.
Each state estimator used 1000 particles.

We see the influence of $\alpha$ that we would expect: higher values lead to more aggressive flagging of measurements as faults, with both true and false positives increasing as $\alpha$ increases.
The state estimation error is represented by the mean absolute percentage error (MAPE), the average of $| \hat{\rho}_{\ell,k} - \rho_{\ell,k}| / \rho_{\ell,k}$ for all $\ell$ and $k$, with $\hat{\rho}_{\ell,k}$ the $\ell$th entry of $\sum_{p=1}^P x_k^p \cdot p(x_k^p | y_k)$, i.e., the mean of the posterior particle filter PDF.
No state estimators that saw faulty data were able to match the MAPE of the faultless estimator.
However, the estimation error decreases with fault labeling error, and the $\alpha$ with the lowest fault labeling error (11.53\% for $\alpha=0.01$) has only a 3\% increase in MAPE over the fault-free case.

\section{Conclusion}
\label{sec:conclusion}
This paper considered a problem where state estimation is desired, but some measurements may be faulty in unknown ways.
Our method for handling this problem takes advantage of a particle filter's structure to perform real-time nonparametric hypothesis tests against the known non-faulty mode.
Now that we have a tool that can pick out faulty sensors, next steps include convergence analysis or inclusion of adaptive control techniques for estimation of unknown values, including the faulty sensor PDF $g_{k_j}^0(\cdot)$ and the optimum $\alpha$ in \eqref{eq:hypothesis_test}.
For example, to estimate the Bernoulli probability of a sensor fault, $\phi_{k_j}$, one could model it as being drawn from a beta distribution (the conjugate prior of the Bernoulli distribution).
Then, as sensors are marked as faulty or not in \eqref{eq:hypothesis_test}, $\phi_{k_j}$'s distribution can be updated based on standard Bayesian inference \cite{robert2007bayesian}.
 
Our discussion was in the context of sensors that are unintentionally faulty, but these methods should also be applicable for robustness to falsified sensors.
If a third party has control of sensors used by a control system, a malicious actor may feed spoofed measurements to the system to purposely manipulate its operation.
%Under our considered method, the lack of a model for faulty, or spoofed, measurements may make it harder for attackers to ``build around'' known attack signatures.
Our fault-model-free method may complement existing defenses that actively search for patterns indicative of spoofed data: perhaps the lack of an explicit model for faulty or spoofed data may make it harder for attacks to be tailored to evade such a detector.
\vspace{-1pt}

%\addtolength{\textheight}{-12cm}   % This command serves to balance the column lengths
                                  % on the last page of the document manually. It shortens
                                  % the textheight of the last page by a suitable amount.
                                  % This command does not take effect until the next page
                                  % so it should come on the page before the last. Make
                                  % sure that you do not shorten the textheight too much.

%%%%%%%%%%%%%%%%%%%%%%%%%%%%%%%%%%%%%%%%%%%%%%%%%%%%%%%%%%%%%%%%%%%%%%%%%%%%%%%%

%%%%%%%%%%%%%%%%%%%%%%%%%%%%%%%%%%%%%%%%%%%%%%%%%%%%%%%%%%%%%%%%%%%%%%%%%%%%%%%%

%%%%%%%%%%%%%%%%%%%%%%%%%%%%%%%%%%%%%%%%%%%%%%%%%%%%%%%%%%%%%%%%%%%%%%%%%%%%%%%%

%%%%%%%%%%%%%%%%%%%%%%%%%%%%%%%%%%%%%%%%%%%%%%%%%%%%%%%%%%%%%%%%%%%%%%%%%%%%%%%%

\bibliographystyle{IEEEtran}
\bibliography{IEEEabrv,bibliography}

\end{document}